# Roman Dodecahedron as dioptron: analysis of freely available data

**Amelia Carolina Sparavigna**
Department of Applied Science and Technology
Politecnico di Torino, C.so Duca degli Abruzzi 24, Torino, Italy

*Recently I have proposed the Roman Dodecahedra as ancient coincidence rangefinders. Here I discuss several data and references freely available on the Web. After analysis, the common features of these artifacts allow to tell that a Roman Dodecahedron was probably a dioptron. The equation of this optical instrument is given. The paper ends with a discussion on the role of the Web and free information for further studies of these objects.*

## 1. Introduction
Recently I discussed a possible use of Roman Dodecahedra as rangefinders [1,2]. My approach to these gallo-roman artifacts of the second and third century was experimental: after preparing a copy of one of them found at Jublains, France [3], and looking through it, I supposed it was a dioptron (instrument to look through, [4]), useful for ballistic to estimate distances. Dioptra were optical instruments, without lenses, used for measuring angles [5].
I based my works [1] and [2] mainly on Wikipedia and references therein [6]. According to Wikipedia, the function of roman dodecahedra is a mystery. "Speculated uses include candlestick holders…; dice; survey instruments; devices for determining the optimal sowing date for winter grain; that they were used to calibrate water pipes; and army standard bases." For what concerns their use as dice, let me remark that a Roman Dodecahedron (see Fig.A1 in Appendix A), which is a small hollow object made of bronze with circular holes of different sizes in the middle of its faces, cannot be used for sorting, because it is a biased body (dodecahedral dice were quite different [7]).
In [1] and [2], I was not able to provide any reference (in English) about the use of Roman dodecahedron for surveying (just a short sentence in a discussion thread [8,9]). Here in the following, I will discuss some references that I found after I published [1] and [2]. These references allow me to conclude that the roman dodecahedron is a dioptron. Moreover, I analyse several of these objects, the data of which are freely available on the Web. The common features of these dodecahedra are providing the equation of the dioptron. The paper ends with a discussion on the role of the Web and free information for further studies of these objects, quite important to understand whether a standard existed or not.

## 2. Dodecahedra for surveying
In previous papers [1,2] I told that the dodecahedral dioptron works as a rangefinder (see Appendix and Figure A1 therein). The equation of the rangefinder is:

$$L = GH \times B/(D_\alpha - D_{\alpha'}), \qquad (1)$$

where L is the distance of a target having size GH, seen through the dodecahedron, fitted in the field of view of the pair (α',α). B is the distance of the two holes. Since the dodecahedron can be considered as a coincidence rangefinder, B is its baseline. $D_\alpha, D_{\alpha'}$ are the diameters of the two opposite holes of pair $(\alpha,\alpha')$.

In [1,2], data of a dodecahedron from Jublains, France, were used: I found these data because the reference was easily available on the Web, under a Google Search in English. Unfortunately, I was not able to find further useful information in English. After a Google Search for images of dodecahedra, I shifted to a search in French (dodécaèdre bouleté) and in German (Pentagondodekaeder), finding some interesting references on the use of roman dodecahedra for surveying. One is Ref.10, a paper in German by Amandus Weiss: it is enough the English abstract and some figures at Ref.11 to understand that the author proposed the use of dodecahedra for surveying as given by Fig.A1. Instead of using it as a rangefinder, he proposed the dodecahedron as a theodolite, with a specific tripod. The equation to evaluate the distance is the same of course. Another quite interesting reference [12] is freely available on the Web. It is again discussing the dodecahedron as a theodolite: in fact it seems that this theory was first proposed [13] "by Friedrich Kurzweil in 1957, of the dodecahedron as a surveying instrument, by which a given distance could be quickly laid out on the ground without the use of tapes" [14,15].

I have not seen the original papers [10,13,14,15], but [11] and [12], freely available, allowed me to known that there were researchers considering the roman dodecahedron as a dioptron: unfortunately, this information was not generally acknowledged by archaeologists. For instance, the authors of Ref.3, consider the dodecahedron as dice for divination; as previously told, these are biased dice, therefore providing just one sort.

## 3. Analysis of some dodecahedra

In this section, let us see in details the data concerning some dodecahedra. To compare the data I use a unit of length u to have diameters $D_\alpha, D_{\alpha'}$ and baseline B as integer multiples of it. In this way, I have the dimensionless ratio $B/(D_\alpha - D_{\alpha'})$ as a ratio of integers (hopefully within the uncertainties of measures, that I do not know, because not reported in literature).

### 3.1 Jublains, France [3]

A dodecahedron having distance B between faces (baselines) ranging from form 48 to 52 mm. The holes are labelled according to Fig.A1. In this dodecahedron, faces 1 and 1' are elliptic, 26×21.5 mm and 25.5×21.5 mm, respectively. In the right column of the table, the baseline is approximated to B = N u, and $|D_\alpha - D_{\alpha'}| = n$ u, with u = 0.5 mm. B is ranging from 48 to 52 mm, then the value of N is slightly varying according to the corresponding pair. Since I have not the precise value for each pair, I suppose N as a constant in this table and in all the following tables.

For the Jublains dodecahedron, B = (100 ±4) u.

| Pair; $(D_{\alpha'}, D_\alpha)$ in mm | $B/(D_\alpha - D_{\alpha'}) = N/n$; u=0.5 mm |
|---|---|
| (1,1') elliptic (26×21.5 mm), (25.5×21.5 mm) | B/(26–25.5)=B/u=N |
| (6',2); (22-21.5) | B/(22–21.5)=B/u=N |
| (5',3); (17-16.5) | B/(17–16.5)=B/u=N |
| (4',4); (22-21) | B/(22–21)=B/2u=N/2 |
| (3',5); (15.5,11.5) | B/(15.5–11.5)=B/8u=N/8 |
| (2',6); (10.5,17) | B/(17–10.5)=B/13u=N/13 |

Table I

### 3.2 Avenches, Suisse [16]

It is a dodecahedron having a diameter of 58.5 cm, and then distance B of 46.5 mm. The holes are labelled according to Fig.A1. $B = (42 \pm ?)$ u.

| Pair; $(D_{\alpha'}, D_{\alpha})$ in mm | $B/(D_{\alpha}-D_{\alpha'}) = N/n$; $u=(1.1\pm0.1)$ mm |
|---|---|
| (6',2); (14.2,15.4) | $B/(15.4-14.2)=B/u=N$ |
| (4',4); (14.5,13.4) | $B/(14.5-13.4)=B/u=N$ |
| (3',5); (8.7,10.4) | $B/(10.4-8.7)=B/(1+1/2)u=N/(1+1/2)$ |
| (5',3); (20.6,18.3) | $B/(20.6-18.3)=B/2u=N/2$ |
| (1,1'); (24.2,26.5) | $B/(26.5-24.2)=B/2u=N/2$ |
| (2',6); (20.2,17.6) | $B/(20.2-17.6)=B/(2+1/2)u=N/(2+1/2)$ |

Table II

Comment. The hole sizes had been measured by Sandrine Bosse Buchanan, Musée romain d'Avenches. The reader can find them in Appendix B, where it is possible to estimate the tolerance of diameters and uncertainties. It seems that the artificer used a tolerance 0.1 mm and a non-decimal system of measures.

### 3.3 Carnuntum and Tongre [12]

| Diameters of pair (mm) | $B/(D_{\alpha}-D_{\alpha'}) = N/n$; $B=40$ mm, $u=0.5$ mm |
|---|---|
| (20.1,20.3) | $B/(20.3-20.1)\approx 2\times B/u=2\times N$ |
| (13.2,13.7) | $B/(13.7-13.2)=B/u=N$ |
| (21.4,22.4) | $B/(22.4-21.4)=B/2u=N/2$ |
| (25,26.5) | $B/(26.5-25)=B/3u=N/3$ |
| (15.3,17.3) | $B/(17.3-15.3)=B/4u=N/4$ |
| (13,10.5) | $B/(13-10.5)=B/5u=N/5$ |

Table III (Carnuntum)

Comment: A unit of u=0.2 mm seems too small. Moreover, no tolerance is given in the reference. Therefore, I prefer to use a unit of 0.5 mm.

| Diameters of pair (mm) | $B/(D_{\alpha}-D_{\alpha'})) = N/n$; $B=63$ mm, $u=0.5$ mm |
|---|---|
| (16,16.2) | $B/(16.2-16)\approx 2\times B/u=2\times N$ |
| (7.5,8.5) | $B/(8.5-7.5)=B/2u=N/2$ |
| (10.5,12.5) | $B/(12.5-10.5)=B/4u=N/4$ |
| (20,22.5) | $B/(22.5-20)=B/5u=N/5$ |
| (12.5,15.5) | $B/(15.5-12.5)=B/6u=N/6$ |
| (16.5,12.5) | $B/(16.5-12.5)=B/8u=N/8$ |

Table IV (Tongres)

Comment: Unit u=0.2 mm seems too small. No tolerance available. I prefer using unit of 0.5 mm.

### 3.4 Vienne (Isère) [18]

Dodecahedron having a diameter of 55 mm, then the baseline is 44 mm.

| Diameters of pair (mm) | $B/(D_\alpha - D_{\alpha'}) = N/n$; B=44 mm, u= 1 mm |
|---|---|
| (24,23) | $B/(24-23)=B/u=N$ |
| (22,19) | $B/(22-19)=B/3u=N/3$ |
| (20,14) | $B/(20-14)=B/6u=N/6$ |
| (20,14) | $B/(20-14)=B/6u=N/6$ |
| (22,15) | $B/(22-15)=B/7u=N/7$ |
| (13.5,22) | $B/(22-13.5)=B/17u=N/(8+1/2)$ |

Table V

### 3.5 Dodecahedron Coulon [19]

| Pair; Diameters of pair | $B/(D_\alpha - D_{\alpha'}) = N/n$; B=?, u=0.25 grid unit |
|---|---|
| (1,1'); (6.5u,6.25u) | $B/(6.5u-6.25u)=N$ |
| (5',3); (4.5u,3.5u) | $B/(4.5u-3.5u)=N/4$ |
| (3',5) (3.5u,2u) | $B/(3.5u-2u)=N/6$ |
| (4',4); (3u,5u) | $B/(5u-2u)=N/8$ |
| (2',6); (2.25u,5.5u) | $B/(5.5u-2.25u)=N/12$ |
| (6',2), (5u,5u) | No angle |

Table VI

The table was obtained using an image from Ref.19 and a grid, to evaluate the diameters of holes. The unit u was assumed equal to ¼ of the grid unit. See Fig.C1 of Appendix C. The labels are the same as in Fig.A1

### 3.6 Dodecahedron Coulon [19]

| Pair; Diameters of pair | $B/(D_\alpha - D_{\alpha'}) = N/n$; B=?, u=0.25 grid unit |
|---|---|
| (3',5) (4u,4.5u) | $B/(4.5u-4u)=N/2$ |
| (1,1'); (6.5u,5.5u) | $B/(6.5u-5.5u)=N/4$ |
| (6',2), (4.25u,3u) | $B/(4.25u-3u)=N/5$ |
| (4',4); (6.5u,5u) | $B/(6.5u-5u)=N/6$ |
| (2',6); (4u,2.5u) | $B/(4u-2.5u)=N/6$ |
| (5',3); (3u,4.75u) | $B/(4.75u-3u)=N/7$ |

Table VII

The table was obtained using an image from Ref.19 and a grid, to evaluate the diameters of holes. The unit u was assumed equal to ¼ of the grid unit. See Fig.C2 of Appendix C.

Unfortunately, in preparing the Tables (besides Table II), I was not able to give all information, in particular about the uncertainties of measures. For this reason, in the case of Tables III and IV, I decided to use u = 0.5 mm, larger that the value 0.2 mm, the minimum difference in diameters I see published in the corresponding reference..

**4. How to figure out**
According to Ref.19, the roman dodecahedra were made by means of a lost-wax casting. This is a quite old technique. As I have discussed in Ref.20, artists of the Bronze Age were able to create quite complex and precise decorations. Therefore, let us assume that the artificer of a dodecahedron was able to have a tolerance of diameters of about 0.2 millimetres. After this assumption, let us discuss how an end-user could figure out a distance from it.
First, we can write the equation of the rangefinder (1) in the rational approximation:

$$L = GH \times B/(D_\alpha - D_{\alpha'}) = GH \times N/n \qquad (2)$$

where n are the integers appearing in the right-columns of Tables I-VII. Integer N depends on the used dodecahedron. The baseline is $B = N\,u$, and the difference between diameters: $|D_\alpha - D_{\alpha'}| = n\,u$. Assuming N and n as fixed values, the measurement uncertainty (in percentage) is given as $\Delta L/L = \Delta(GH)/GH$, where $\Delta(GH)$ is the uncertainty of GH. Let us assume the estimate of target dimensions to be about 10%. Therefore, the uncertainty after a single measure of distance was about 10%.
From the Tables, it seems that there was a constant unit u of about one or one-half of millimetres. However, we must remember that Romans had a non-metric and non-decimal system of measurement [21], and then, this unit of one or one-half mm that we find in the analysed of some dodecahedra is amazing. Since there are more than a hundred of dodecahedra in museums and collections, it would be quite important to have a statistics of this unit to decide whether a standard measure existed or not. In particular, it is necessary to provide for each dodecahedron a table as that given in Appendix B (dodecahedron of Avenches).
For what concerns a non-decimal system, this seems used in Tables I, II and V.
Probably, the artificer provided a table reporting the values of integers N and n, accompanying the dodecahedron. In the case that these data were unknown, anybody could measure them using compass or caliper (the roman "circinus", [22]): an outside caliper for the baseline B and a divider caliper for the holes. The outside caliper is used to mark the two ends of the baseline B on a wax tablet. The divider caliper measures and marks the diameter D of the larger hole on the tablet, fitting a tip in one end of B and the other drawing an arc (red in Figure 1). Then the caliper is used to measure D' of the smaller hole. Fitting a tip where the red arc crosses the baseline, the second (violet) arc is outlined. The difference is $|D - D'| = n\,u$.

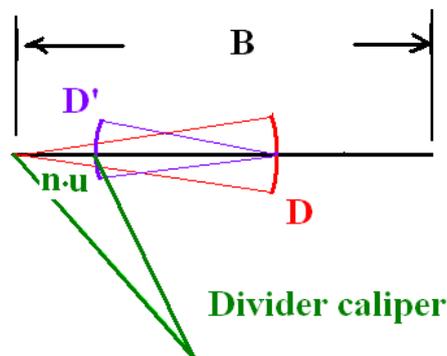

**Fig.1** We can use the "circinus" to measure the baseline B and a divider caliper for the holes. The circinus is used to mark the two ends of the baseline B on a wax tablet. The divider caliper marks the diameter D of the larger hole, fitting a tip in one end of B and the other drawing an arc (red). Then the caliper is used to measure D' of the smaller hole. Fitting a tip where the red arc crosses the baseline, the second (violet) arc is outlined. The difference is $|D - D'| = n\,u$.

It is easy to obtain the ratio N/n, turning the caliper enough times on B. Let us call the number of turns T. In this case, instead of (2), we have:

$$L = GH \times T \qquad (3)$$

The uncertainty (percentage) is $\Delta L/L = \Delta(GH)/GH + \Delta T/T$. Of course, T and $\Delta T/T$, can be deduced after a calibration of the instrument. In this case, too, an approximate uncertainty for a single measure of about 10% can be used, according to the uncertainty of the target estimation.

Let us note that for writing, and then for calculation too, Romans used wax tablets, which were tablets covered of wax. Then, the fact that wax was found inside a dodecahedron does not mean that it was a candlestick holder [6], but simply that wax was used too by the owner of dodecahedron.

We have also to remark that the faces of the dodecahedra are often decorated with concentric engraved circles. These grooves are quite good for measures using calipers, because the tips fit in them.

The dodecahedron of Jublians has a decoration composed of three circles. Each pair of opposite holes of the artifact of Jublains has the same decoration, as shown in the following image (Fig.2) arranged from an image of Ref.3. This could be also a mnemonic code for calculations.

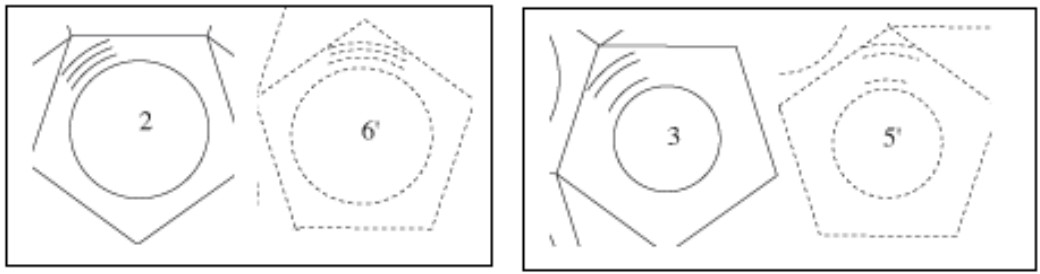

**Fig.2. Engraved circles on the faces of Jublains (from an image of Ref.3).**

It could be interesting to use caliper and measure the diameters of these grooves. May be, they are related to diameters of holes or to the equation of rangefinder (2) and (3). Without further information, any conclusion about these grooves is impossible.

In some of the Tables of Section 3, we see that there are pairs of holes having the same ratio N/n. This does not mean that the precision of the measure that they are giving it the same. Let us remember that the measurement of distance is based on the length GH of the target seen through the holes. Changing the pair of holes can improve the precision of the measure, because the target can have a better visibility. For what concerns the knobs on the dodecahedra, they were quite useful for a good grip of them.

**5. Discussion and conclusions**

In this paper I have analyses some dodecahedra, the data of which I found freely available on the Web (except 3.2, the data were kindly sent by e-mail). The common features of these artifacts allow concluding that a Roman Dodecahedron was probably a dioptron. This optical instrument has an equation that can be written in a form (2), which is including a ratio of integers. As proposed by [10-14], the dodecahedron was used as an ancient theodolite, placed on a tripod and with a range-pole as a target - may be painted as the modern ones - for precise measurements.

It does not seem that a standard or rule for these instruments existed. We see from the tables a certain recurrence of u equal to 1 or one-half of millimetres, approximately, but it is better to collect more data before reaching any conclusion that has a statistical significance. For this analysis, the role of the Web is fundamental, in particular if free information about these objects is given.

Unfortunately, as I have directly experienced, there are some biases. For instance, the studies on the use of dodecahedra as surveying instruments are quite old, but it seems that archaeologists still do not acknowledge them. For this reasons, several hypotheses on their use are prospering on the Web, concealing the relevant ones (Rari nantes in gurgite vasto). Here again, the role of the Web is crucial in increasing the knowledge and further interactions between scientific disciplines.

**6. Acknowledgments**
Many thanks to **Sandrine Bosse Buchanan, Musée Romain d'Avenches (Avenches)** for the data of the dodecahedron.

**Appendix A**
The use of a dodecahedron as a rangefinder is shown in Fig.A1 On the left we see a bronze dodecahedron. The diameter was usually ranging from 4 to 11 centimetres. On the right, we see the faces of a dodecahedron found at Jublains [3]. In the lower part of the image, an angle of view for the distance measurement is shown. The angle of view is the cone that describes the angular extent of a given scene. Knowing the size of the scene we can determine its distance [1,2].

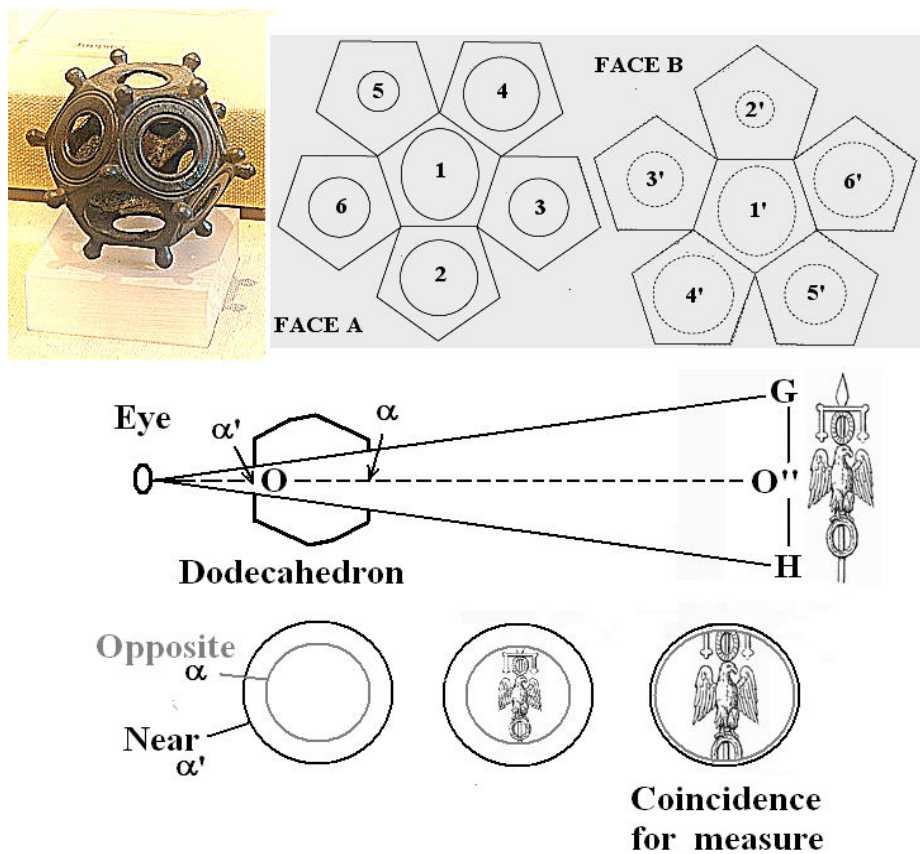

**Fig.A1 On the left a dodecahedron (source: Wikipedia). On the right, we see the faces of the roman dodecahedron of Ref. 3. In the lower part of the image, we can see how the dodecahedron is used. Let us consider a pair of opposite hole (α', α), for instance (2', 6), and look at a Roman vexillum through the dodecahedron, holding it with α' and α parallel, and α' near an eye and α opposite. If the dodecahedron is close the eye, we see the two holes. If it is too far, we can see just hole α'. There is a distance where we can see the circumferences of the two holes (black and grey in the image) as perfectly superimposed. This coincidence is giving a specific cone of view that we can use for measurements. For this reason, we could tell that the dodecahedron is a coincidence rangefinder [2].**

Points O, O'', G and H are given in Fig.A1. OO'' is the eye-line. Let us imagine a Roman soldier looking through the dodecahedron at a target of which he knows approximately its size (GH). If the soldier sees the target fitted in the field of view of the pair ($\alpha'$, $\alpha$), that is, with GH as diameter of $\alpha'$ and $\alpha$ in coincidence, he can evaluate the distance OO'' from the following relation: Let us call B the distance of the two holes. From similar triangles we have: OO''=B×OO'/($D_\alpha$−$D_{\alpha'}$), where ($D_\alpha$−$D_{\alpha'}$), are the diameters of the two holes ($\alpha$,$\alpha'$). Of course, it is not necessary that the target is perfectly fitting the hole. It is necessary that the soldier, looking through this dioptron, is able to estimate the size of the scene seen through it. If we consider the decahedron a rangefinder, B is its baseline.

**Appendix B**
The hole sizes had been measured by Sandrine Bosse Buchanan, Musée romain d'Avenches.
The data I received are given in the following table (left column). Where two values are given, these are the smaller and the larger diameter measured for the hole.

| Diameters (smaller and larger) in mm | Approximation and uncertainties |
|---|---|
| 1:  26.46, 26.62 | 26.5±0.1 |
| 2:  15.44 | 15.4 |
| 3:  18.31 | 18.3 |
| 4:  13.4 | 13.4 |
| 5:  10.43 | 10.4 |
| 6:  17.38, 17.85 | 17.6±0.2 |
| 1':  24.16, 25.08 | Probably elliptic: 24.2, 25,1 |
| 2':  19.9, 20.43 | 20.15±0.2 |
| 3':  8.68 | 8.7 |
| 4':  14.27, 14.68 | 14.5±0.2 |
| 5':  20.4,  20.83 | 20.6±0.2 |
| 6':   14.18 | 14.2 |

Table B1

In Table II, the values from the right column of Table B1 had been used.
It seems that the artificer that created the dodecahedron found at Avenches used a tolerance lower than 0.2 mm in a non-decimal system of measure.

**Appendix C**

Reference 19 is giving some images from a book written by R. Coulon, dated 1910. Ref.19 is telling that "C'est précisément ce qui pousse R.Coulon à conclure que ces objets doivent être envisagés comme des exercices de maîtres, des sortes de chefs-d'œuvre. La plupart des auteurs s'accordent à reconnaître que les dodécaèdres ont été coulés selon la technique de la cire perdue."

An image of the site [19] allowed me to evaluate the data from two dodecahedra, as shown in Fig.C1 and C2. A grid was superimposed to the image from the book. The grid helps in evaluating the diameters of holes. The unit u was assumed equal to ¼ of the grid unit. In the following figures, the diameters are shown in red and the labels of faces in black.

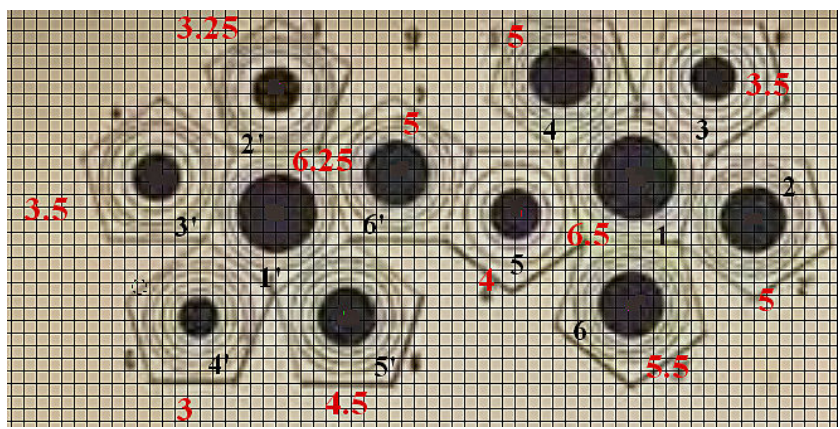

**Fig. C1**

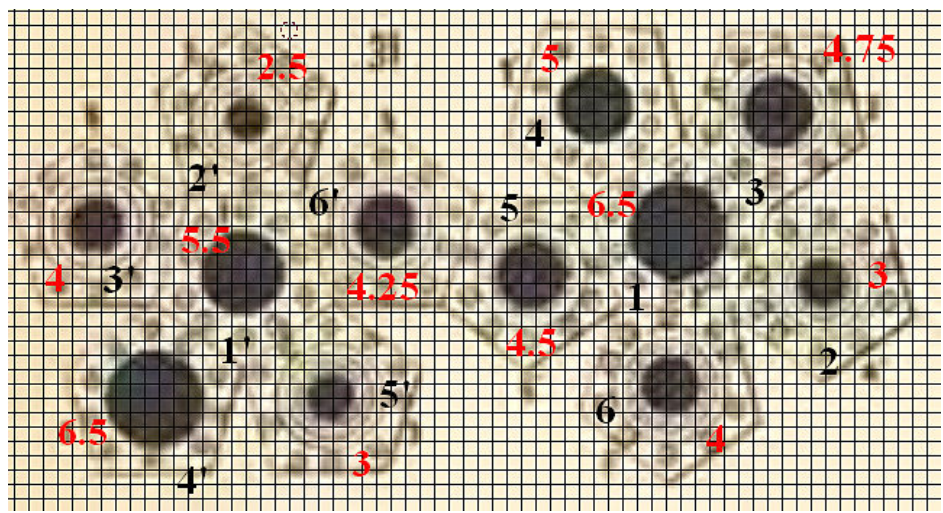

**Fig. C2**

**References**
1. A.C. Sparavigna, A Roman Dodecahedron for measuring distance, 2012, arXiv, arXiv:1204.6497v1 [physics.pop-ph], http://arxiv.org/abs/1204.6497
2. A.C. Sparavigna, Ancient and modern rangefinders, 2012, arXiv, arXiv:1205.2078v1 [physics.pop-ph], http://arxiv.org/abs/1205.2078
3. G. Guillier, R. Delage and P.A. Besombes, Une fouille en bordure des thermes de Jublains (Mayenne) : enfin un dodécaèdre en contexte archéologique!, Revue archéologique de l'Ouest, Vol.25, p. 269-289, 2008
4. From Guglielmo Gemoll, 1959, Vocabolario Greco-Italiano; διοπτευω, osservare, considerare da tutti i lati, esplorare (to observe, consider all sides, explore); διοπτηρ, esploratore (ranger); διοπτρα (Polibio, X, 46,1), livella, traguardo, diottra (diopter), διοπτρον, istrumento per guardar attraverso (instrument to look through), διορατι, atto di vedere attraverso (looking through); διοραω, guardare attraverso, discernere (to look through, distinguish).
5. M.J.I. Lewis, Surveying Instruments of Greece and Roe, Cambridge University Press, 2001.
6. http://en.wikipedia.org/wiki/Roman_dodecahedron, retrieved 2 June, 2012
7. A.C. Sparavigna. An Etruscan Dodecahedron, 2012, arXiv, arXiv:1205.0706v1 [math.HO], http://arxiv.org/abs/1205.0706
8. After reading several discussion threads on this subject, I have found, at http://sculpture.net/community/showthread.php?t=10801, what SteveW wrote in 2011: "This object is an on-the-fly surveying device. Worn on a chain around the neck, a military officer could hold it out to arms length and "fit" an enemy soldier or catapult, ballista etc between the opposing holes and then multiply by 100 (or whatever factor/number of links of the chain) to gauge his distance. It would have been accurate enough to place a man within a few feet up to roughly a mile away."
9. The references on roman surveying methods are usually discussing, besides the "dioptra", the "groma" and the "surveyor's cross", another dioptron. See for instance: Cesare Rossi, Flavio Russo, and Ferruccio Russo, Ancient Engineers' Inventions: Precursors of the Present, Springer, 2009; Wilhelm A. Schmidt, The Surveyor's Cross, Professional Surveyor Magazine, March 2007, http://www.profsurv.com/magazine/article.aspx?i=1803).
10. Amandus Weiss, Zu den Anwendungsmöglichkeiten des Pentagonododekaeders bei den Römern [Possible uses of Roman 'dodecahedra'], Archäologisches Korrespondenzblatt, 5, 1975, 221-4. In http://www.biab.ac.uk/contents/32027 we can find the abstract: "Returns to earlier theories for the use of the Roman dodecahedron as a measuring device, as opposed to the candle-stick explanation (see 71/1018). Three possibilities are suggested: as a measurement gauge, with the holes providing a continuous scale of half-scrupulus units; as a surveying instrument for setting out a prescribed distance on the ground or determining the distance between two points; and thirdly (a new suggestion), as a planning instrument for setting out horizontal angles in towns and buildings, with a subdivision into new degrees (g) of 1/400 as opposed to conventional degrees of 1/360. F H T."
11. http://www.felixweiss.ch/nazcadocs.htm and links therein, retrieved 2 June, 2012.
12. Pierre Méreaux - Tanguy, Le Dodecaedre:: Mesureur d'Angle?, Kadath, May-June July, 1975. The author is telling that "Les ouvertures percées en regard dans deux faces opposées semblent identiques à l'œil, mais, après vérification par mesure, on constate une légère différence entre elles. Ceci avait déjà attiré l'attention de l'ingénieur Friedrich Kurzweil, en 1956, lors d'un examen détaillé du dodécaèdre conservé au Museum Carnuntinum, en Autriche. Grâce à l'obligeance de Monsieur Smeesters, directeur du Musée gallo-romain de Tongres, j'ai pu relever les mesures exactes du dodécaèdre de ce musée et les résultats, assez étonnants, confirment ceux obtenus à l'aide de celui de Carnuntum."

13. F.H. Thompson, Dodecahedrons Again, The Antiquaries Journal, Volume 50, Issue 01, 1970, pp 93-96. The author is telling that "K. Mauel discussed and elaborated a theory first proposed by. Friedrich Kurzweil in 1957, of the dodecahedron as a surveying instrument, by which a given distance could be quickly laid out on the ground without the use of tapes."
14. F. Kurtweil. Das Pentagondodekaeder des Museum Carnuntium und seine Zweckbestimmung. in Carnuntum Jarbuch, 1956, pp.23-29.
15. K Mauel, Der 'Theodolit' des römischen Feldmessers. in Vereins Deutscher Ingenieure Nachrichten, 29 nov, 1961, p.14.
16. Avhenches, Sandrine Bosse Buchanan, Musée romain d'Avenches, 2012, e-mail communication of data on the dodecahedron of Aventicum.
17. http://www.aventicum.org/de/Musee/expop/documents/MRA_angl_web.pdf
18. Duval Paul-Marie. Comment décrire les dodécaèdres gallo-romains, en vue d'une étude comparée. In: Gallia. Tome 39, fascicule 2, 1981. pp. 195-200, doi : 10.3406/galia.1981.1829 http://www.persee.fr/web/revues/home/prescript/article/galia_0016-4119_1981_num_39_2_1829
19. Les dodécaèdres gallo-romains ajourés et bouletés, discussion on the paper by R. COULON, Essai de reconstitution des dodécaèdres creux, ajourés et perlé. Rouen-Caignard, 1910, http://letarot.com/pages-vrac/pages/Truc-a-trous.html
20. A.C. Sparavigna, Number pi from the decoration of the Langstrup plate, 2012, arXiv, arXiv:1203.4103v1 [physics.pop-ph], http://arxiv.org/abs/1203.4103
21. A.C. Sparavigna, Cooking the volumes, arXiv, 2012, arXiv:1201.3774v2 [physics.pop-ph], http://arxiv.org/abs/1201.3774
22. On the CIRCINUS, compass, see the link
http://penelope.uchicago.edu/Thayer/E/Roman/Texts/secondary/SMIGRA*/Circinus.html